\documentclass[12pt,preprint2]{emulateapj}

\include{epsf}
\usepackage{graphicx}
\usepackage{times}
\newcommand{\hi}{H\,{\sc i}}
\newcommand{\htwo}{{H}$_{2}$}

\newcommand{\hubble}{\mbox{$\rm km\, s^{-1}\, Mpc^{-1}$}}
\newcommand{\kms}{\mbox{$\rm km\, s^{-1}$}}

\newcommand{\mhi}{\mbox{$M_{\rm HI}$}}

\newcommand{\nhi}{\mbox{$N_{\rm HI}$}}

\newcommand{\icmsq}{\mbox{$ \rm cm^{-2}$}}
\newcommand{\fnhi}{\mbox{$f(N_{\rm HI})$}}

\newcommand{\lya}{Ly$\alpha$}
\newcommand{\dv}{\mbox{$\Delta V_{90}$}}

\slugcomment{\aj,  in press}
\shorttitle{Kinematics of DLAs}
\shortauthors{Zwaan et al.}

\begin{document}

\title{Are the kinematics of DLAs in agreement with their arising in the gas disks of galaxies?}

\author{Martin Zwaan\altaffilmark{1}\email{mzwaan@eso.org}}
\author{Fabian Walter\altaffilmark{2}}
\author{Emma Ryan-Weber\altaffilmark{3}}
\author{Elias Brinks\altaffilmark{4}}
\author{W.J.G. de Blok\altaffilmark{5}}
\author{Robert C. Kennicutt, Jr\altaffilmark{3}}


\altaffiltext{1}{ESO, Karl-Schwarzschild-Str. 2, Garching 85748, Germany}
\altaffiltext{2}{Max-Planck-Institut f{\"u}r Astronomie, D-69117 Heidelberg, Germany}
\altaffiltext{3}{Institute of Astronomy, University of Cambridge, Madingley
     	Road, Cambridge CB3 0HA, UK}
\altaffiltext{4}{Centre for Astrophysics Research, University of Hertfordshire, Hatfield 
	AL10 9AB, UK}
\altaffiltext{5}{Department of Astronomy, University of Cape Town, Private Bag X3, 
	Rondebosch 7701, Republic of South Africa}

\begin{abstract}
We demonstrate in this paper that the velocity widths of the neutral
gas in Damped Ly$\alpha$ (DLA) systems are inconsistent with these
systems originating in gas disks of galaxies similar to those seen in the
local universe. We examine the gas kinematics of local galaxies using
the high quality \hi\ 21-cm data from the \hi\ Nearby Galaxies Survey
(THINGS) and make a comparison with the velocity profiles measured in
the low-ionization metal lines observed in DLAs at high redshifts. The
median velocity width of $z=0$ \hi\ gas above the DLA column density
limit of $\nhi=2\times 10^{20}\,\icmsq$ is approximately 30\,\kms,
whereas the typical value in DLAs is a factor of two higher. We argue
that the gas kinematics at higher redshifts are increasingly influenced
by gas that is not participating in ordered rotation in cold disks, but is more likely
associated with tidal gas related to galaxy interactions or processes such as superwinds and outflows.
An analysis of the \hi\ in the local interacting star-burst galaxy M82 shows that the velocity widths in this
galaxy are indeed similar to what is seen in DLAs.

\end{abstract}

\keywords{galaxies: ISM --- 
ISM: evolution --- 
ISM: kinematics and dynamics ---
radio lines: ISM ---
galaxies: statistics ---
quasars: absorption lines}

%###############
\section{Introduction}
Until the next generation of large radio telescopes 
systematically detects the 21-cm hyperfine line of neutral, atomic hydrogen (\hi)
in emission at redshifts beyond $z>0.2$, our knowledge of the \hi\ at high
redshift will remain solely dependent on the study of absorption-line
spectra of high redshift quasars. These spectra are dominated by
\lya\ absorption lines originating in intervening \hi\ clouds, and show
a large range in \hi\ column density. The systems with the largest column densities, termed
Damped \lya\ Absorbers \citep[DLAs;][]{Wolfe1986a},
are characterized by $\nhi>2 \times 10^{20}\,\icmsq$ and are known to hold
most of the \hi\ atoms in the universe, at all redshifts
\citep[cf.][]{Peroux2005a,Zwaan2005a}. A popular explanation is that
the DLAs are the progenitors of gas disks in present-day galaxies \citep{Wolfe2005a}.

At $z=0$ it is well established that most of the \hi\ atoms are locked
up in the gas disks of $L_*$-type galaxies \citep{Zwaan2003a}. These
$z=0$ \hi\ disks show an frequency distribution of \hi\ column densities [\fnhi] that is
virtually identical in shape to that of the DLAs
\citep[][]{Ryan-Weber2003a, Zwaan2005a,Prochaska2005a}. Furthermore,
at redshifts $z<1$ there are approximately 20 identifications of
galaxies giving rise to DLA absorption in spectra of background quasars
\citep{LeBrun1997a, Steidel1995a, Turnshek2001a, Chen2003a, Rao2003a,
  Lacy2003a}. The properties of these galaxies (such as luminosities and
impact parameters) are fully consistent with
the idea that the DLAs arise in galaxies that are typical in the local universe \citep{Zwaan2005a}. Also the metallicities
seen in low redshift DLAs are fully consistent with what is expected
when gas disks of $z=0$ galaxies are intercepted randomly. Therefore,
DLAs at low redshifts ($z<1$) probably arise in gas disks of galaxies like
those seen in the local universe.

At higher redshifts ($z>1$) the situation becomes more
complicated. Direct detection of DLA host galaxies has proven to be
extremely difficult \citep[see e.g.,][and references
  therein]{Moller2002a,Moller2004a}. This may not be
surprising if at these epochs the DLAs also arise in gas disks of
normal spiral galaxies: the cross-section selection ensures that a
large fraction of the DLA galaxies is fainter than $L_*$, and thus difficult to detect
\citep{Fynbo1999a,Zwaan2005a}. The shape of the column density
distribution function, \fnhi, at these redshifts is still consistent with what is seen at
$z=0$. Although this is consistent with the galaxy disk
model, it is inconclusive evidence for this model. It is conceivable that
the shape of \fnhi\ is determined by rather simple physics, for example ionization at low 
column densities and \htwo\ formation at high densities. This might produce the same \fnhi\ at all redshifts,
regardless of whether this gas is distributed in ordered disks or
not.

Other observations of high-$z$ DLAs are also inconclusive
\citep[see][for a comprehensive review]{Wolfe2005a}. The inferred cosmic
star formation rate density from DLAs is consistent with that measured
from Lyman Break Galaxies \citep{Wolfe2003a}. However, metallicities
of the DLAs are typically low \citep[e.g.,][]{Pettini1999a}, as are the molecular
fractions \citep[e.g.,][]{Ledoux2003a}, and analyses of
elemental abundances show star formation histories typical of those of
quiescent galaxies \citep{Dessauges2006a}.

Perhaps one of the most conclusive tests of whether DLAs at high redshift
arise in gas disks or not, can be made by studying their
kinematics. The first work on this topic was
presented by \cite{Prochaska1997a}, who compiled high resolution, high
signal-to-noise spectra of low-ionisation metal lines in DLAs, and compared
them with the velocity profiles derived from simulations using model ensembles of artifical clouds with a range of velocities. They concluded that rapidly rotating thick
disks are the most likely hosts of DLAs. This conclusion was contested
by \citet{Haehnelt1998a}, who showed that irregular protogalactic
clumps can reproduce the absorption line properties of DLAs equally
well. A direct comparison with 21-cm data of local galaxies was
first made by \citet{Prochaska2002b}, who found that low mass galaxies such as
the Large Magellanic Cloud (LMC) display much lower velocity widths than
normally seen in DLAs.

A fair comparison with DLA kinematics can only be made by
studying a representative sample of local galaxies for which high quality 21-cm
data cubes exist, and which have high spatial and frequency
resolution and low \hi\ column density limits. In this paper we make
use of the THINGS \citep[The \hi\ Nearby Galaxy Survey;][]{Walter2008a} data, which are
ideal for this purpose. 

It is important to note that the \lya\ lines that define the DLA
carry little information on the kinematics of the absorbing systems themselves. The velocity
width that would correspond to the total extent of the damping wings is much
larger than the physical velocity width corresponding to the
turbulence or large scale movements of the gas; any structure in the
velocity distribution of the \hi\ gas is completely washed out by the
damping wings. Instead, workers in this field rely on low-ionization metal lines, such as
Fe$^+$, Si$^+$, Ni$^+$, which are believed to be good tracers of the neutral gas
\citep[e.g.,][]{Prochaska1997a}. In the same system each of these
transitions gives consistent velocity widths, as long as the lines are
unsaturated. In principle, many small clouds in the outskirts of galaxy halos
could contribute to the total velocity profile measured in these low-ionization lines.
However, the parameter that we use to characterize the velocity width, 
\dv\,  (the velocity width that encompasses the central 90\% of the total optical depth
seen in the line) ensures that these low column density clouds do not
contribute to the total width.

 In section \ref{approach.sec} we describe our strategy and 
 data, in section \ref{results.sec} we present the results, and in
 section \ref{interpretation.sec} we discuss these results. Finally, we present the conclusions
 section \ref{conclusions.sec}. Throughout this
 paper we use a value for the Hubble constant of $H_0=75\,\hubble$, although the distances to some of 
 the THINGS galaxies are derived using direct distance measurements.

%###################
\section{The approach}
\label{approach.sec}
For a quantitative comparison between the kinematics of DLAs and local
galaxies, high quality 21-cm maps are required that are characterized
by high signal to noise and high spatial and velocity resolution. A
low noise level is required to sample all regions of the gas disks
above the DLA column density limit of $\nhi=2\times 10^{20}
\icmsq$. A high spatial resolution is needed to avoid severe
overestimation of the velocity profiles due to spatial smoothing
\citep[see][]{Ryan-Weber2005a}. Since for aperture synthesis
interferometry the noise level (measured as the minimal detectable \hi\ column density 
per beam) scales with
the inverse of the beam area, these two requirements are very
difficult to meet. The THINGS sample is currently the only sample with
sufficient spatial resolution and depth, and contains a large enough
number of galaxies such that the local
universe galaxy population can be sampled representatively.

\subsection{THINGS}
\label{things.sec}
We base our analysis on VLA observations of a sample of nearby galaxies 
drawn from THINGS: The HI Nearby Galaxy Survey \citep{Walter2008a}. 
THINGS provides high angular and velocity resolution data
of uniform quality for 34 galaxies spanning a range of 
Hubble types, from dwarf irregular galaxies to massive spirals. The sample 
selection is mainly drawn from the Spitzer Infrared Nearby Galaxy Survey \citep[SINGS,][]{Kennicutt2003a} but excludes targets with distances $>$15 Mpc, and 
elliptical galaxies. In addition, edge--on galaxies have 
been excluded. The observations and data reduction are discussed in detail 
in \citet{Walter2008a}. For our analysis, we use 
the robust-weighted data cubes, which produce typical beam sizes of $6"$
(or 200\,pc for the mean distance of 6.5 Mpc to our sample galaxies). 
We choose the robust-weighted data cubes in order to achieve the highest possible 
spatial resolution, which is required for making a comparison to DLA velocity
profiles, which are measured along very narrow sight-lines to background quasars. 
The velocity resolution in all cases is equal or better than 5.2 \kms\ (which we will smooth further, 
see below). Typical column density sensitivities are 
$5\times10^{19}\,\icmsq$ in the integrated \hi\ maps/profiles, i.e., much 
below the DLA column density limit ($2\times 10^{20}\,\icmsq$) discussed in this paper.

The different distances among the galaxies in the sample cause the beam to trace 
different physical scales within galaxies. As discussed in section \ref{beam.sec}, beam
smearing causes some increase in the measured velocity widths. Therefore, for the most 
distant galaxies the velocity width measurements are increased somewhat more  than for the nearby ones. However, as we will see, this does not affect our main conclusions.

\subsection{Biases in the THINGS sample}
\label{bias.sec}
The THINGS sample is composed of galaxies from optical catalogues, 
whereas DLA systems are purely
\hi-selected. \citep[see][for a discussion on the selection of THINGS
  targets]{Walter2008a}. However, in \citet{Zwaan2005a} it was argued
that this optical selection does not lead to any biases for this application. The main
argument is the following: \citet{Zwaan2000c} has shown that the
properties of \hi-selected galaxies are not different from those of
optically selected galaxies if compared at the same
luminosity. Therefore, if the weighting of the sample is properly
taken into account (using the luminosity function or \hi\ mass
function) no strong biases should be introduced by using optically
selected galaxies.  This fact is also corroborated by the calculation
of \citet{Ryan-Weber2003a}, who showed that the redshift number
density $dN/dz$ calculated from \hi-selected nearby galaxies is
consistent with the values presented in \citet{Zwaan2005a} and in the
present paper.

Another consideration is the distribution of inclinations $i$ of
the THINGS sample. Ideally, the galaxies should be randomly oriented,
that is, the distribution of $\cos i$ should be consistent with a flat
distribution. In practice, however, given the selection discussed in section~\ref{things.sec},
the sample has a minor deficiency of edge-on
galaxies. A Kolmogorov-Smirnov test shows that the null-hypothesis that
the inclination distribution is consistent with being random can only
be rejected at the 19\% level.  Because of projection effects, the
most inclined galaxies have \hi\ column densities and \dv\ measurements at the 
high end of the total distribution of these values in the THINGS sample.
On the other hand, the
cross-section in highly inclined galaxies is very small, meaning that
the fractional contribution of these galaxies to the total count of
high column densities and high \dv\ values is minor.  If we assume that
THINGS misses all galaxies with inclinations $i$ greater than $80^{\circ}$, this
would lead to a total incompleteness in the number of galaxies of $\cos 10^\circ=17$\% (assuming infinitely thin disks). However, since 
the area $A$ of a disk scales as $A\sim \cos i$, the total area missed in these
highly inclined galaxies is only $\int_{80}^{90} \cos i \sin i \,di / \int_{0}^{90} \cos i \sin i \,di \approx$ 3\%. We conclude that the contribution from the missing highly inclined galaxies 
to the DLA cross-section is not a significant concern for this analysis.

\subsection{Beam Smearing}
\label{beam.sec}

If the kinematics of the \hi\ change on spatial scales smaller than
the beam size, measurement of the gas properties are subject to
systematic uncertainties. This effect is known as beam
smearing. \citet{Ryan-Weber2005a} demonstrate that decreasing the
spatial resolution of 21-cm data systematically increases the measured
median \dv. This result was established by convolving a high
resolution \hi\ data cube of the Large Magellanic Cloud (LMC) with
Gaussian beams of various widths to simulate data with different
spatial resolutions. The increase in \dv\ with beam width can be
understood in terms of including adjacent pockets of gas that are
offset in velocity. The resulting velocity width measurement of the
total area will be greater than the \dv\ values of the individual
\hi\ clouds. The LMC data has a physical resolution of 15 pc beam$^{-1}$
and a velocity resolution of 1.6 \kms. In comparison, the mean
physical resolution of THINGS is 200 pc. Given this factor of $\sim15$
difference in physical resolution, and employing the same technique
used by \citet{Ryan-Weber2005a}, we estimate that the THINGS velocity
width measurements could be overstated by $\approx$ 40\%. Despite this 
expectation of an increase, the velocity dispersion
and measurements from the LMC and THINGS data are similar.
Specifically, the distribution of velocity dispersion values in the
LMC peaks at 6 \kms, wheras values derived from the THINGS galaxies
range from 6 to 10 \kms, i.e. not larger in all lines-of-sight. Thus,
we conclude that beam smearing may have a mild effect on the THINGS
data, but is unlikely to cause a major systematic shift in the
measured \dv\ values. Of course the physical scales sampled by the
highest resolution 21-cm measurements are still larger than those
probed by QSO lines-of-sight. In any case, as we will see in Section \ref{results.sec}, 
if we have overestimated the velocity widths of galaxies at $z=0$ using the
THINGS data this would imply an even stronger discrepancy between the true low redshift velocity widths and the high redshift  \dv\ values of in DLAs.

\subsection{The column density distribution function}
We have argued that THINGS provides the most suitable sample of 21-cm data
cubes of nearby galaxies to perform the kinematic analysis. However,
the number of galaxies in this sample is an order of magnitude smaller
than the WHISP\footnote{The Westerbork \hi\ survey of Spiral and
Irregular galaxies \citep{Hulst2001a} is an \hi\
survey of 355 galaxies carried out with the Westerbork Synthesis
Radio Telescope. } sample used in \citet{Zwaan2005a}, which provides a very accurate
measurement of the local \hi\ column density distribution function
\fnhi. 

To test whether the THINGS sample contains biases in
the \hi\ column densities it samples, we first calculate the \hi\ column density distribution function \fnhi. This
function is defined such that $f(\nhi) d\nhi dX$ is the number of
systems with column density between $\nhi$ and $\nhi+d\nhi$ over
a distance interval $dX$. We
use the method described by \citet{Zwaan2005a}, where \fnhi\ is
calculated as follows:
 \begin{equation}
\fnhi=\frac{c}{H_0} \frac{\sum_i \Theta(\mhi_i) w(\mhi_i) A_i (\log
  \nhi)}{\nhi \, \ln 10 \,\Delta \log \nhi }.
\label{fn.eq}
\end{equation}
 Here, $\Theta(\mhi_i)$ is the space density of galaxy $i$, measured
 via the \hi\ mass function. For the \hi\ mass function we adopt
 the most recent HIPASS measurement as published by
 \citet{Zwaan2005b}.  The function $w(\mhi_i)$ is a weighting function
 that takes into account the varying number of galaxies across the
 full range of $\mhi$, and is calculated by taking the reciprocal of
 the number of galaxies in the range $\log \mhi_i-\Delta/2$ to
 $\log \mhi_i+\Delta/2$, where $\Delta$ is taken in this case to be 0.35 dex.  $A_i(\log
 \nhi$) is the area function that describes for galaxy $i$ the area in
 $\rm Mpc^{2}$ corresponding to a column density in the range $\log
 \nhi$ to $\log \nhi+\Delta \log \nhi$.
In practice, this area function is simply calculated by summing for each galaxy 
the number of pixels in a certain $\log \nhi$ range multiplied by the physical
area of a pixel. 
Finally, $c/H_0$ converts the number of systems per Mpc to
that per unit redshift.

\begin{figure}
\begin{center}
\includegraphics[width=7.8cm,trim=0cm 0cm 0cm 0cm]{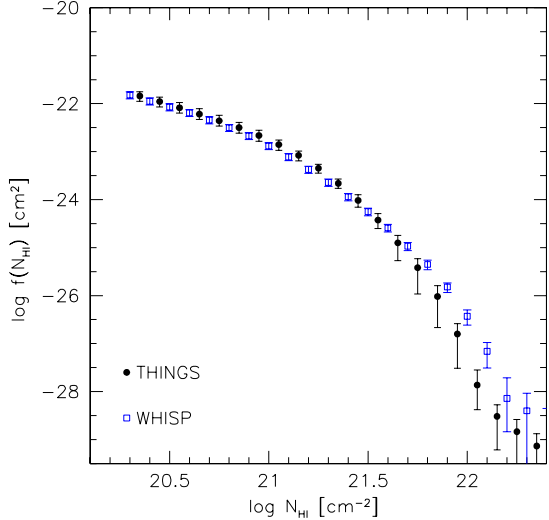}
\caption{The \hi\ column density distribution function at $z=0$ from
  21-cm emission-line observations. The open symbols represent the
  calculation of \citet{Zwaan2005a} based on the WHISP sample, whereas
  the solid circles represent the calculation presented in this paper
  based on THINGS data. Error bars include uncertainties in the
  \hi\ mass function as well as counting statistics and indicate
  1$\sigma$ uncertainties.
\label{fnthings.fig}}
\end{center}
\end{figure}

Figure~\ref{fnthings.fig} shows the resulting column density
distribution function \fnhi\ measured from the THINGS data (solid
points), compared to the one published by \citet{Zwaan2005a} based on
the WSRT maps of the WHISP sample (open points). Given the different 
sample definitions of THINGS  and WHISP and  the different resolutions achieved in both surveys,
there is a remarkable agreement between the two data sets, reaffirming our earlier conclusion that
selection effects are not an important issue for the THINGS sample,
and thus that this sample is well suited for this kind of analysis. Note
that we only calculated the \fnhi\ points above the DLA column density
limit of $\nhi=2\times10^{20}\,\icmsq$ (corresponding to $\log \nhi=20.3$), since this is the limit we apply for our
kinematic analyses. In
principle, the THINGS \hi\ column density sensitivity would allow us
to calculate \fnhi\ down to much lower \nhi\ values.

Some differences between the two samples can be seen at high column
densities, above $\log \nhi\approx 21.7$. This is due to the
dearth of high inclination galaxies in THINGS. Although the effect shows up 
clearly in the logarithmic representation of \fnhi, we showed in section \ref{bias.sec} that the
effect of missing edge-on galaxies on the total \hi\  cross section and on the \dv\ 
calculation is not important for this analysis.

\begin{figure*}
\begin{center}
\includegraphics[angle=0,width=\textwidth]{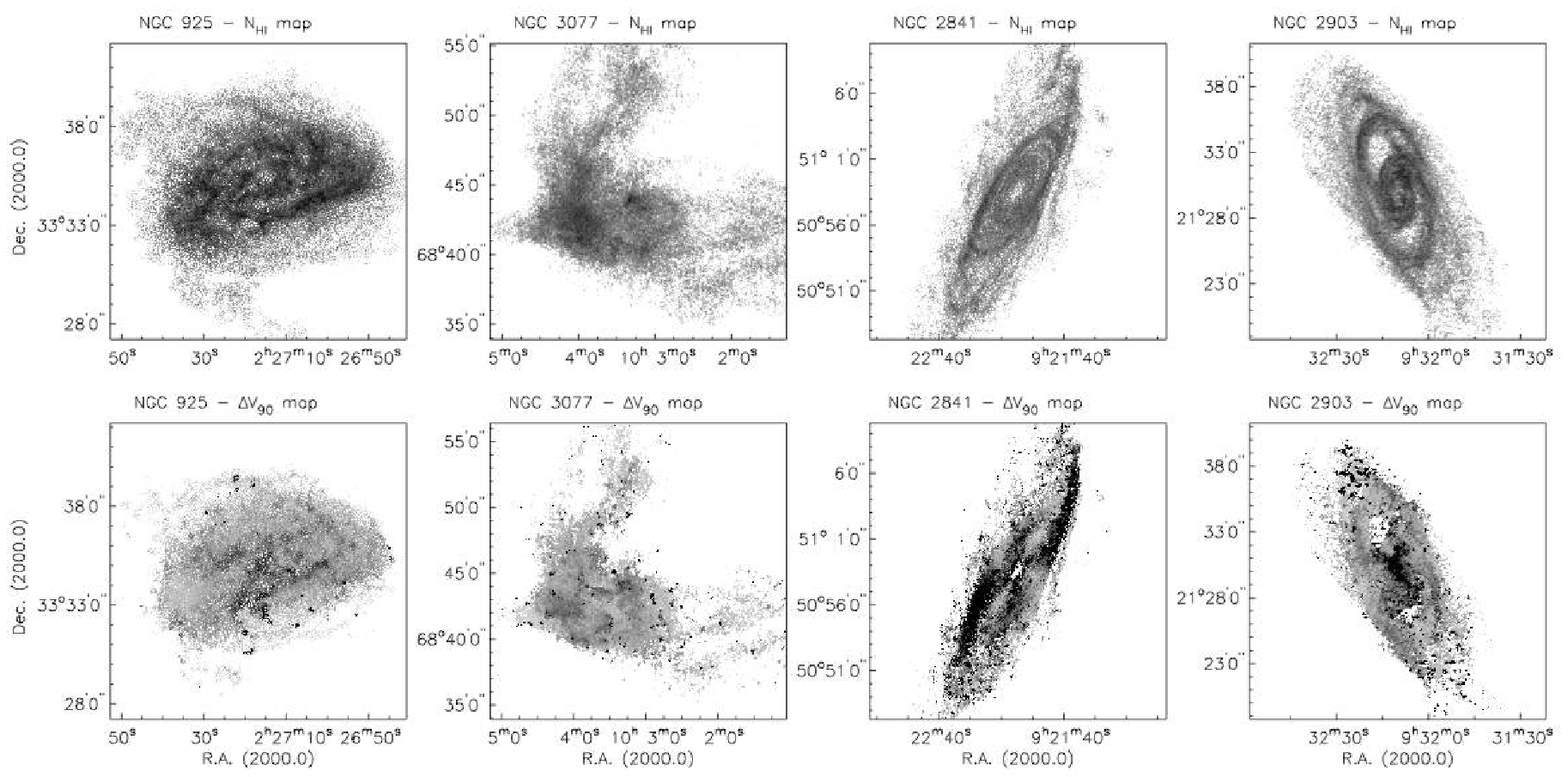}
\caption{Examples of \hi\ maps and \dv\ calculations for four of the THINGS galaxies: NGC 925, NGC 3077, NGC 2841, and NGC 2903. The top row of panels shows the \hi\ column density distributions. The greyscales range from $\log\nhi=$ 19.8 to 21.8.
The bottom row of panels shows
the distribution of \dv\ over the region of the galaxies for which the \hi\ column density exceeds the DLA limit of $2\times 10^{20}\,\icmsq$. The full range of the greyscale is 0 to 100 \kms. The black regions are those where \dv\ exceeds 100 \kms. The black spots that can be seen for example in the northern and southern outskirts of NGC\,2903 are due to noisy spectra, where the \dv\ calculation is unreliable. In NGC\,2841, the extended dark regions in the \dv\ map are associated with high \nhi\ regions and indicate true high values of \dv.
\label{mosaic.fig}}
\end{center}
\end{figure*}

The zeroth moment of \fnhi\ gives the redshift number density $dN/dz$
of gas above the DLA \hi\ column density threshold in the $z=0$ universe. This quantity simply 
represents the number of encounters with \hi\ gas above the DLA limit that a 
hypothetical observer would have if they were traveling through the local universe
for a total path length of $\Delta z=1$.
 From the
THINGS sample we find $dN/dz=0.047 \pm 0.009$, compared to
$dN/dz=0.045\pm0.006$ from the WHISP analysis, which again is in
excellent agreement.

\subsection{The velocity widths}
The standard metric used in DLA studies to characterize the profile
width of metal absorption lines is \dv\,, the velocity
width that encompasses the central 90\% of the total optical depth
seen in the line. This definition assures that weak velocity components
that are not good tracers of the large-scale velocity field, as well as
statistical fluctuations, are not taken into account.  In practice,
\dv\ is calculated by taking the cumulative optical depth
distribution, and finding the velocities at which this distribution
crosses the 5\% and 95\% levels. In DLA studies, the data used for
this analysis are normally limited to those with high signal to noise,
where the peak optical depth exceeds $20\sigma$. These spectra are
mostly products of the High Resolution Echelle Spectrometer (HIRES) on the Keck telescope 
and the Ultraviolet and Visual Echelle Spectrograph (UVES) on the Very Large Telescope (VLT). 
This signal-to-noise level is calculated on smoothed
spectra with typical velocity resolutions of 15 to 20 \kms.

In order to make an equitable comparison, we first smooth the THINGS data
cubes in velocity with a boxcar filter, to a velocity resolution of
approximately 15 to 20 \kms, varying slightly from cube to cube,
depending on the original channel width. Next, we calculate normalized
cumulative velocity profiles at every spatial pixel in the cube. Using
the same strategy as defined above (just cutting the distribution at
the 5\% and 95\% levels) does not give good results, because for the lower
column densities the \hi\ data typically possess lower S/N than the 
optical spectra (of which only the highest S/N examples are regarded). 
Therefore, the cumulative distribution can
cross these levels several times, leaving us with no unique definition
of \dv. To avoid this problem, we first
identify the 50\% level and travel outward in both directions to find
the first instance where the 5\% and 95\% levels are crossed. The
difference between the two velocities at which this happens is then
defined as \dv. In principle, this method could cause an underestimation of 
\dv, but tests with synthetic spectra show that this effect is minimal.

Figure \ref{mosaic.fig} shows four examples of the \dv\ calculations for the
THINGS galaxies. We choose to show the nearly face-on barred spiral 
NGC\,925, the M81-group member NGC\,3077 know for its extended tidal HI, the isolated 
nearly edge-on early-type spiral NGC\,2841, and the barred nuclear starburst galaxy NGC\,2903. 
The top row of panels shows the \hi\ column density distribution for these galaxies. 
The bottom row of panels shows the distribution of \dv\ over the regions
of these galaxies where the \hi\ column densities exceed the DLA threshold of 
$\nhi=2\times 10^{20}\,\icmsq$. The black regions are those where \dv\ exceeds 100 \kms. The black spots that can be seen for example in the northern and southern outskirts of NGC\,2903 are due to noisy spectra, where the \dv\ calculation is unreliable. In NGC\,2841, the extended dark regions in the \dv\ map are associated with high \nhi\ regions and indicate true high values of \dv.

\begin{figure}
\begin{center}
\includegraphics[width=7.8cm,trim=2cm 0cm 1.6cm 0cm]{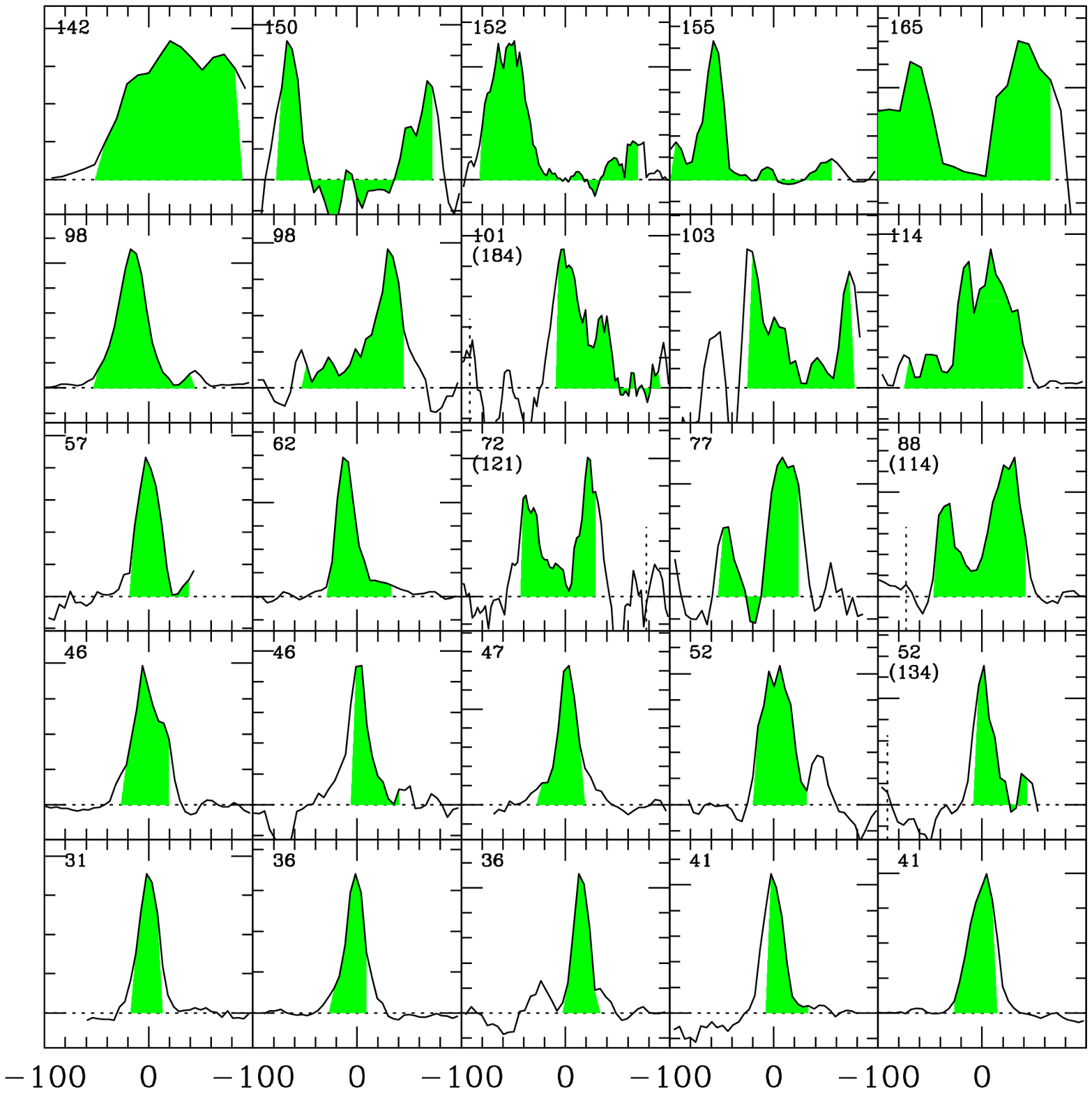}
\caption{Representative \hi\ profiles from the THINGS data. The profiles are sorted 
from bottom to top by increasing \dv\ values. The shaded areas correspond to the inner 90\% of the optical depth and define the measurement of \dv. The number in the upper left corner of each panel shows the value of \dv\ for each profile, whereas the number in parenthesis shows \dv\ based on second crossings of the 5\% and 95\% levels (see text). The dashed vertical lines show the velocities at which the profiles reach these second crossings.
The vertical scaling is arbitrary, the horizontal axis runs from $-100$ to $100$ \kms\ for all profiles.
\label{profiles.fig}}
\end{center}
\end{figure}

Examples of velocity  profiles through THINGS galaxies are shown in Figure \ref{profiles.fig}.
We chose 25 random profiles, with \dv\ values between 31 and 165 \kms.
This figure gives an idea of quality of the THINGS data and the accuracy of the \dv\
measurements. The shaded areas correspond to the inner 90\% of the optical depth and define the measurement of \dv. The number in the upper left corner of each panel shows the value of \dv\ for each profile. To indicate the reliability of our method for determining \dv\ by using the first crossing of the 5\% and 95\% levels from the center, we also show in parenthesis the values of \dv\ based on the second crossing of these levels. For most profiles, the cumulative distribution crosses the 5\% and 95\% levels only once, in which case the definition of \dv\ is unique. In the few cases where the profiles are more noisy and a second crossing occurs, it is clear that the \dv\ measurements based on the first crossings are superior; measurements based on the second crossing clearly overestimate \dv. These example profiles show that our method of determining \dv\ is accurate and not introducing a significant bias.

\subsection{Do low ionization metal lines trace only the neutral gas?} \label{lowion.sec}
Before presenting the results, we would like to come back to the statement made in the introduction that low ionization metal lines such as Fe$^+$, Si$^+$, Ni$^+$ trace the neutral gas only. The fundamental premise of this paper is that these lines trace the kinematics of the high column density neutral gas, and can therefore be compared directly to local 21-cm measurements. 

It is well established that approximately 25\% of a spiral galaxy's hydrogen mass is in a layer of ionized gas, known as the diffuse ionized gas \citep[DIG, e.g.][]{Reynolds1993a}. The scale height of this gas is 0.9 kpc in the Milky Way Galaxy, much larger than that of the neutral gas seen in 21-cm. If a large fraction of the optical depth seen in low-ionization lines observed in DLAs would arise in such a medium, our results might be biased because this medium does not contribute to the profile of the \hi\ 21-cm line.

\citet{Sembach2000a} present model calculations of ionization corrections in the DIG of the Milky Way, and conclude that elemental abundances in ionized gases might be biased without significant  ionization corrections. However, much more relevant in the context of the present paper are studies directed at  DLAs specifically since our concern is whether the ionized gas influences the low-ionization metal line measurements of DLA profiles. A main motivation for such studies of ionization effects in DLAs is the observation that Al$^{+2}$ is present in DLAs with the same radial velocity distribution as that of low ionization lines \citep{Lu1996a}. However, the results of e.g., \citet{Viegas1995a}, \citet{Vladilo2001a} and \citet{Lopez2002a} show conclusively that ionization effect corrections for low-ionization lines observed in DLAs are negligible. The direct implication of this is that the velocity width measured in low-ionization lines is not affected significantly by an ionized medium. For systems below the DLA \hi\ column density limit, the situation might be very different \citep{Dessauges-zavadsky2003b}, but these are not considered in the present analysis.

%#############
\section{Results}
\label{results.sec}
We first use the THINGS data to calculate the probability distribution function of the
velocity width \dv\ of galaxies responsible for high column density
\hi\ absorption at $z=0$. This function defines the number of systems
above a certain column density limit that would be encountered along a
random 1 Mpc path through the $z=0$ universe, as a function of \dv.
The function is calculated by multiplying the cross sectional area of \hi\ as a function of 
\dv\ by the space density of galaxies. If DLAs at $z=0$ arise in galaxies similar to those observed in the local universe,
this function would describe the expected distribution of \dv\ values for DLAs.
For the details of calculating the probability
distribution function we refer to \citet{Zwaan2005a}.

\begin{figure}
\begin{center}
\includegraphics[width=7.8cm,trim=1cm 6cm 5.5cm 0cm]{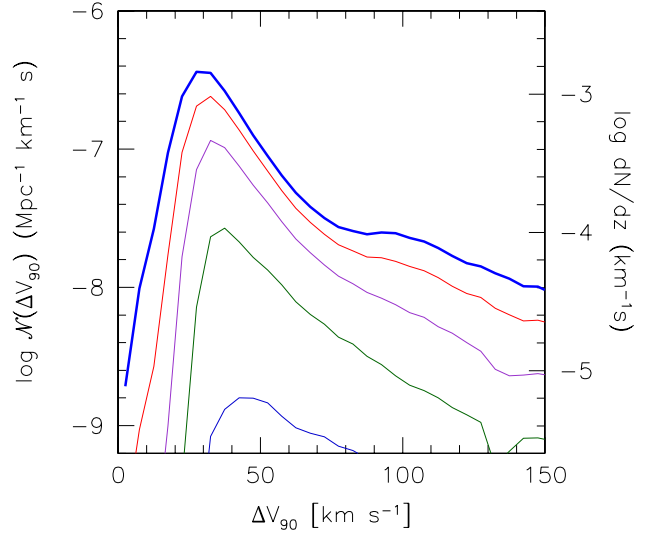}
\caption{The distribution of velocity widths \dv\ of high
  column density gas $z=0$ from the THINGS data. The lines show the 
  number of systems as a function of \dv\ encountered along a 
  random 1\,Mpc sightline. The right axis
  shows the corresponding number of systems per unit redshift
  $dN/dz$. Under the hypothesis that DLAs arise in gas disks similar to those
  seen in the local universe, these curves show the expected \dv\ distributions for DLA absorbers.
  The different lines correspond to different column density
  limits: the thick, uppermost curve corresponds to
  the classical DLA limit of $\log \nhi>20.3$, the other curves are for 
  $\log \nhi>$ 20.6, 20.9, 21.2, 21.5, respectively.
\label{W90dens.fig}}
\end{center}
\end{figure}

Figure~\ref{W90dens.fig} shows the result. The lines show the
distributions above column density limits of $\log \nhi=$ 20.3, 20.6,
20.9, 21.2, and 21.5, from top to bottom, respectively. For all column densities above the
DLA limit of $\log \nhi=20.3$ the distribution peaks at $\sim 30$ \kms,
and shows a long  tail toward higher \dv\ values. As one progresses to higher
column density cut-offs, the peak of the distribution shifts toward larger
\dv\ values, but the upper tail remains. We will discuss this further
in section \ref{dv-nhi.sec}.

\begin{figure}
\begin{center}
\includegraphics[width=5.8cm,trim=2cm 0cm 2cm 0cm]{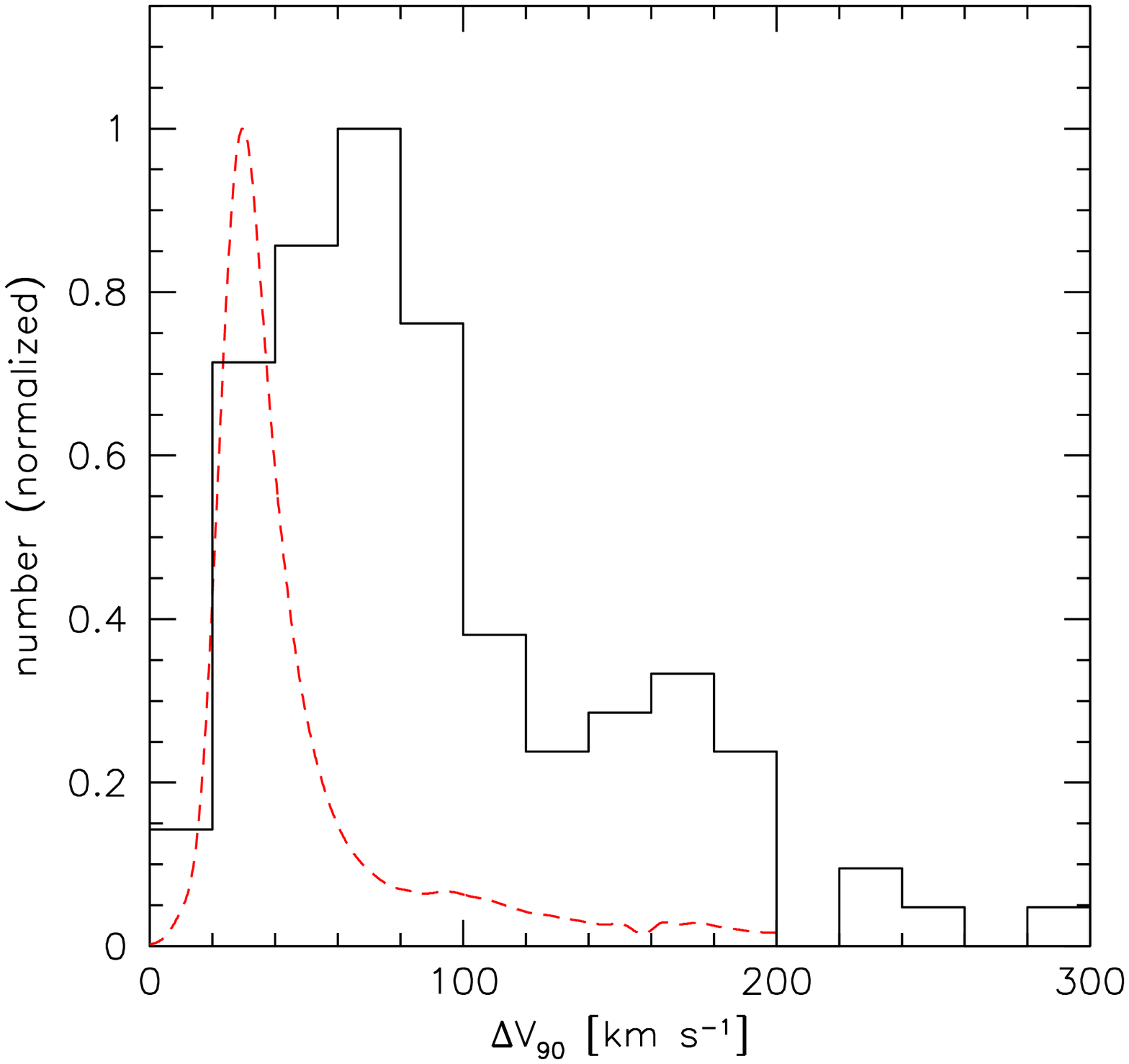}
\caption{The distribution of \dv\ measured in local galaxies (dashed line) and in
  DLAs (histogram). The DLA data are taken from \citet{Prochaska2007a}. The local galaxy data are based 
  on THINGS and only include sightlines where the \hi\ column density exceeds the 
  DLA limit of $\nhi=2\times 10^{20}$ \icmsq. The
  galaxy data are weighted such that they represent cross-section
  selected gas. The dashed line is the 
  same as the top solid line in Figure~\ref{W90dens.fig}, but now on a linear scale.
\label{histV90.fig}}
\end{center}
\end{figure}

Immediately obvious from this distribution is that the peak in \dv\ is
much lower than what is typically seen in DLAs at higher
redshifts. This is illustrated directly in Figure~\ref{histV90.fig},
where we show on a linear scale the total probability distribution
function of \dv\ for $z=0$ gas above the DLA limit (as a dashed line),
together with the measured distribution of \dv\ in DLAs at redshifts
between $z=1.5$ and $z=4$ from \citet{Prochaska2007a}. Note that the dashed line is the same as the 
top solid line in Figure~\ref{W90dens.fig}, but now on a linear scale. 
The shape of the two distributions
is very similar -- a sharp peak with a high \dv\ tail -- but the
location of the peak is approximately a factor two higher for the DLAs
than for the local galaxies based on THINGS.

This result reaffirms the conclusions of \citet{Prochaska2002b}, which showed that the
gas kinematics in local galaxies are inconsistent with the majority of
DLA sightlines. \citet{Prochaska2002b} concentrated on the
gas kinematics of the Large Magellenic Cloud and we have shown that the
same conclusion holds if a more representative sample of local
galaxies is considered. Whereas the kinematics of \hi\ gas above the
DLA limit at $z=0$ is dominated by ordered rotation in cold disks, the velocity
fields of DLA systems at high redshifts must be influenced by other
physical processes such as outflows, superwinds, and interactions. We discuss this in
more detail in section \ref{conclusions.sec}.

\subsection{The relation between \dv\ and \hi\ column density}
\label{dv-nhi.sec}

\begin{figure}
\begin{center}
\includegraphics[width=7.8cm,trim=1cm 0cm 0cm 0cm]{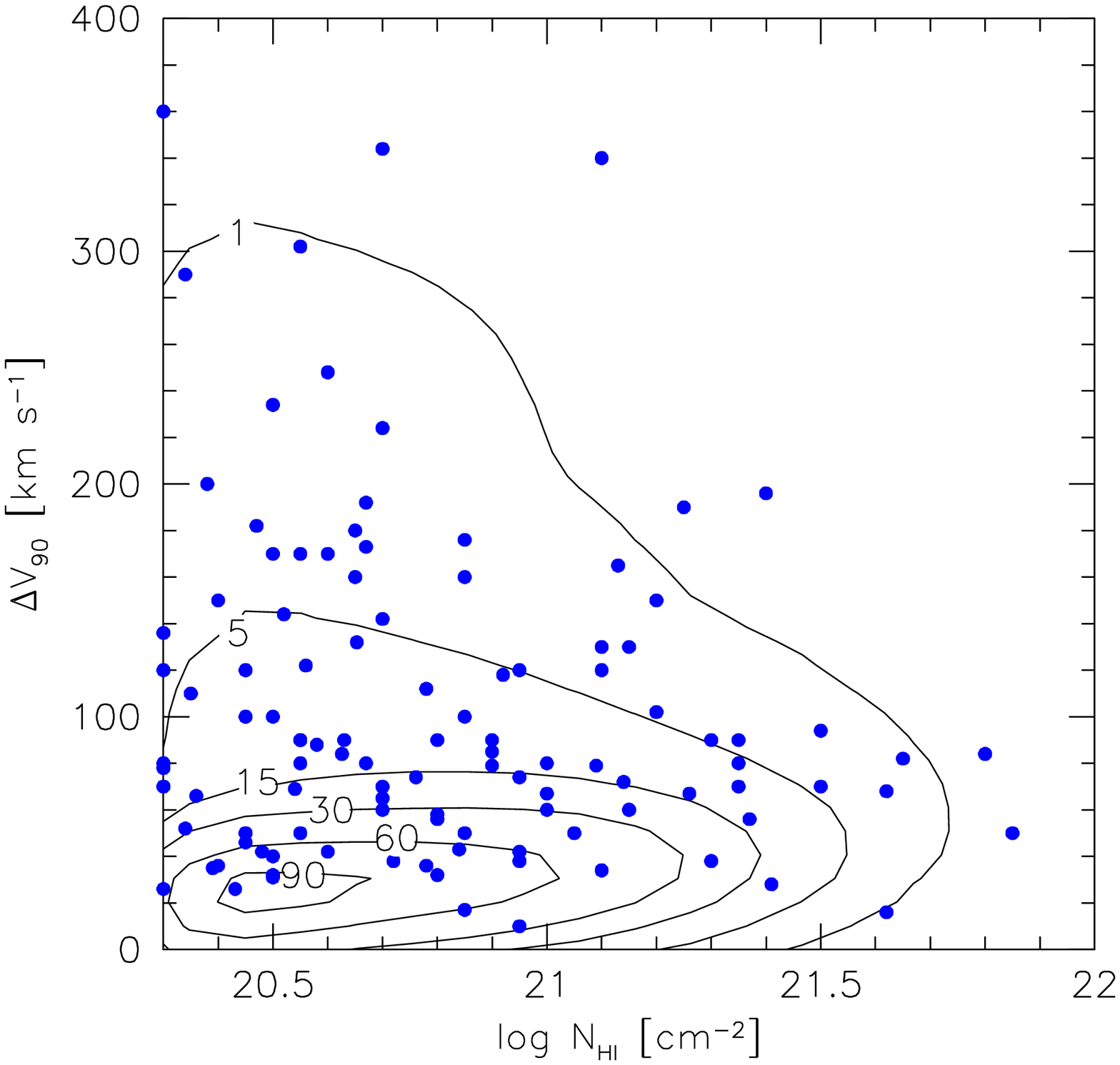}
\caption{The bivariate distribution of \dv\ and \hi\ column
  density. Contours are calculated from the THINGS data, and are drawn
  at probability levels of 1, 5, 15, 30, 60, and 90\%.  The points
  are the DLA data from \citet{Prochaska2007a}. Two effects are
  obvious from this plot: 1) the peak of the $z=0$ distribution is at \dv\ $
  \approx 30\, \rm km\, s^{-1}$, and increases slowly toward larger
  \hi\ column densities; 2) large values of \dv\ occur mostly in
  regions of galaxies with relatively low \hi\ column densities.
\label{W-N.fig}}
\end{center}
\end{figure}

It is interesting to investigate how the velocity spread of the
\hi\ gas depends on the \hi\ column density range considered. The contours in Figure
\ref{W-N.fig} show the probability distribution of \hi\ cross section
in the \nhi-\dv\ plane. The THINGS data were used to calculate the cross sectional area 
contributed by each element $d\nhi d\dv$ on a fine grid in the $\nhi$-$\dv$ plane.
The \hi\ mass
function has again been used to assign weights to individual galaxies. 
The weighting scheme is similar to that used for the calculation of \fnhi,
which is expressed in Equation~\ref{fn.eq}. The points in this figure are the 
DLA data from \citet{Prochaska2007a}, over the redshift range $z=1.5$ to $z=4$.

There are
two notable effects visible in this diagram. First, the peak of the $z=0$
distribution occurs at values of \dv\ around 30\,\kms, and increases
slowly toward larger \hi\ column densities, as already seen from Figure \ref{W90dens.fig}. 
For Gaussian profiles, the 
value of \dv=30
\kms\ translates to a velocity dispersion of $\approx 9$ \kms, which
is a typical value for the gas disks in spiral galaxies. Observationally the 
line-of-sight velocity dispersion ranges from $\approx 6$ \kms\ in the outer parts  to $\approx 12$ \kms\ in the center \citep[e.g.,][]{Meurer1996a,DeBlok2006a,Fraternali2002a}. From the THINGS data it is found that the 
dispersions in the regions away from obvious \hi\ shells is $11 \pm 3$ \kms, averaged over the galaxy disks \citep{Leroy2008a}.
The increase
of \dv\ toward larger \nhi\ values is consistent with the observation
that the \hi\ velocity dispersion in gas disks typically decreases
with galactocentric radius \citep[e.g.,][]{Dickey1990a,Kamphuis1993a}, and the
fact that radial \hi\ column density profiles usually show a decline
toward larger radii \citep{Cayatte1994a, Bigiel2008a}.

The other notable feature in Figure \ref{W-N.fig} is that the larger values of \dv\ are mostly
associated with the lower column densities. To understand this, we
need to investigate further the origin of the high-end tail in the
\dv\ distribution. There are several components contributing to this
tail. One of the most obvious contributions is that of pixels near the
centers of galaxies, where the gradient in the rotation curve is
large. A line of sight through these regions is likely to encounter
gas over a large range in velocity. The steepest gradients are seen in
earlier type spiral galaxies \citep[e.g.,][]{Noordermeer2007a}, where typically the \hi\ densities are
depressed in the center because of the conversion of \hi\ to molecular hydrogen
\citep[e.g.,][]{Wong2002a}. These factors
combined cause a large contribution of low \nhi, high \dv\ points.

\begin{figure}
\begin{center}
\includegraphics[width=7.8cm,trim=0cm 0cm 0cm 0cm]{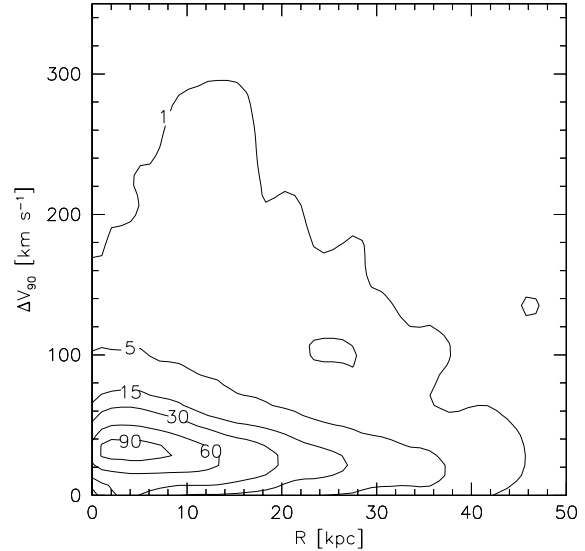}
\caption{The bivariate distribution of \dv\ and $R$, the position from the 
  center of the galaxy. Contours are calculated from the THINGS data, and are drawn
  at probability levels of 1, 5, 15, 30, 60, and 90\%.  Larger values of \dv\ 
  are more likely to occur at small distances from the center of a galaxy.
  Please note that the small scale structure in the 1\% contour is due to noise.
\label{W-R.fig}}
\end{center}
\end{figure}

To illustrate that the larger values of \dv\ are more likely to occur near the
centers of galaxies we constructed Figure \ref{W-R.fig}, which shows the 
probability distribution of \hi\ cross section
in the $R$-\dv\ plane, where $R$ is the distance from the center of the galaxy. It
is important to note that this diagram is also made in such a way that it represents
probabilities for $R$-\dv\ pairs to occur for random sight-lines through the local 
universe. Therefore, $R$ can be regarded as an ``impact parameter", similar to that defined 
in QSO absorption line studies as the projected distance between the position of a
background quasar and that of the center of a galaxy giving rise to an absorption line. 
Clearly, there is an upper envelope in the distribution: the maximum \dv\ seen at any value of the impact parameter $R$ decreases toward larger $R$.

%Another contribution to the high-\dv\ tail comes from tidal \hi. {\bf true?} {\em i would be suprised if the high values were to come from small warps or tidal hi - the beam is so much more narrow than these tidal features. -- I'm making some maps for illustration.}

The simple fact that the lower \nhi\ pixels show noisier
spectra causes another important contribution of high \dv\ points. In lower signal-to-noise 
spectra, noise peaks  more easily contribute to the central 90\% flux contained in the
spectra. This fact can be seen in Figure \ref{profiles.fig}, where 
we showed random velocity profiles through THINGS galaxies. For the lowest 
\dv\ values, the profile is clearly dominated by a single component. For 
$\dv\sim 70$ \kms, the profile typically shows two components, or one strong
component with a lower flux tail. This continues to be true for even higher \dv\
profiles, but here irregular profiles seem to get increasingly important. 

Comparing the contours and the points in Figure \ref{W-N.fig}, we note that there
is some indication for the DLA data to show the same trend as the THINGS 
local galaxies data: higher values of \dv\ are more likely to occur at lower \hi\
column densities.

\subsection{Qualitative comparison of DLA and galaxy velocity profiles}
\label{qualitative.sec}
So far, our analysis has concentrated on the parameter \dv, but it is 
interesting also to consider the shape of the profiles in more detail. 
In the previous section we noted that for sightlines where $\dv\gtrsim 70$ \kms, 
the profile typically shows two components. In fact, it appears that the individual
line components have widths within the main peak of the \dv\ distribution in Figure
\ref{W90dens.fig}, and the high -\dv\ tail is mostly due to line splitting and multiple components. 

Interestingly, this same behavior is seen in the metal absorption line profiles 
of DLAs. For example, Fig 1 of \citet{Prochaska1997a} shows 17 examples of velocity profiles of low-ionization transitions from DLAs. The narrow profiles are characterized by a single component with $\dv\sim 30\,\kms$, very similar to what we observe in the local universe for most of the sightlines through galaxy disks (cf. Figure~\ref{histV90.fig}). For DLA profiles with larger widths, a shoulder or tail becomes progressively more important, and a pronounced second peak starts to separate from the main component for \dv\ values exceeding $\sim 100\,\kms$. Apparently,  the second peak in Figure\,\ref{histV90.fig} beyond $\dv\sim 100\,\kms$ is mostly due to profiles clearly showing multiple components, whereas the main peak corresponds mostly to profiles displaying a single component with a weaker tail. The profiles with multiple components are seen at $z=0$ as well (see Figure\,\ref{profiles.fig}), but are very rare. When we consider just the profiles that correspond to the main peaks in  Figure\,\ref{histV90.fig}, we see that at $z=0$ they are mainly symmetric, single component profiles, whereas the DLAs at higher redshift typically show a tail. This characteristic is important  
for understanding the mechanisms repsonsible for shaping the profiles and will be discussed more in section \ref{interpretation.sec}.

\subsection{Is \dv\ a good estimator of galaxy mass?}
\label{mass.sec}
There is accumulating evidence for the existence of a relation between
DLA (and sub-DLA) velocity spread and metallicity at high redshifts
\citep{Moller2004b,Ledoux2006a, Murphy2007a, Prochaska2007a}. The interpretation of
this correlation, assuming
that \dv\ is correlated with mass, is that the well-known luminosity-metallicity
relation observed in local galaxies \citep[e.g.,][]{Tremonti2004a} was
already in place at redshifts $z=4$ all the way down to $z=2$. As a proxy for luminosity (or
galaxy mass), \citet{Ledoux2006a} use measurements of \dv. The authors
argue that \dv\ measured in low ionization lines is dominated by
motions governed by gravity, irrespective of whether the gas is
following the rotation of the galaxy or related to infall or outflow
of gas or to merging of galaxy sub-clumps. In principle, a large
collection of sightlines through the gaseous regions of galaxies
should show a relation between depth of the potential well and \dv.

\begin{figure}
\begin{center}
\includegraphics[width=7.8cm,trim=0cm 0cm 0cm 0cm]{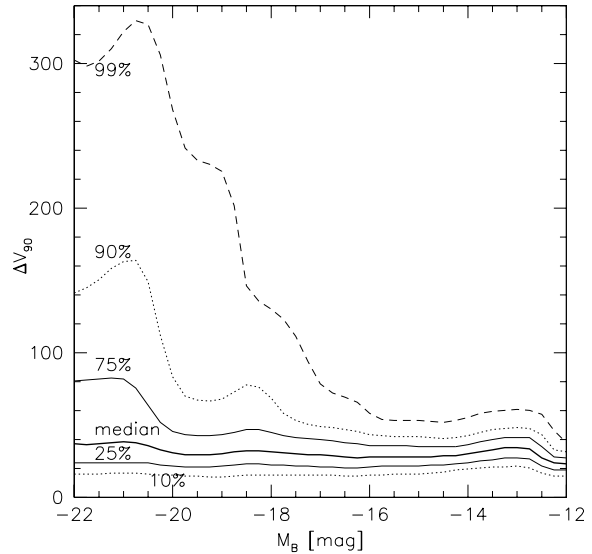}
\caption{The conditional probability distribution of \dv\ versus absolute $B$-band
magnitude, $M_B$. The thick solid line represents the median of the velocity width \dv\ distribution at each value of $M_B$.  The thin solid lines show the 25th and 75th percentile distribution, the dotted lines show the 10th en 90th percentile, and the dashed line shows the 99th percentile distribution. The curves indicate global trends, small scale structure in the curves is due to small number statistics. The median \dv\  remains constant around $\dv=30$ \kms. 
\label{condprobVM.fig}}
\end{center}
\end{figure}

Only at $z=0$ are we in the situation that we can test whether the
assumption that \dv\ is a statistical measurement of galaxy mass or
luminosity is valid.  The result is shown in
Figure~\ref{condprobVM.fig}. Here we show the conditional probability
distribution of \dv\ as a function of absolute $B$-band
magnitude. This figure a created by first calculating the bivariate distribution
of cross-section in the $M_{\rm B}$,\dv\ plane, and then normalizing at each $M_{\rm B}$
 the distribution function of cross-section as a function of \dv.

Obviously, the likelihood of finding a high \dv\ value goes up only marginally toward brighter galaxies: the median does not rise significantly from low luminosity to high luminosity, but for
more luminous galaxies the high \dv\ tail becomes more pronounced. 
Therefore, in a statistical sense, \dv\ is a very weak proxy for host galaxy mass.
At any absolute magnitude, the median value of \dv\ is $\approx 30 \,\kms$, which for a Gaussian profile translates to a velocity dispersion of 9 \kms.

In conclusion, at $z=0$ the velocity profile measured along a thin
line of sight at a random position through a galaxy gas disk is a
very weak indicator of the galaxy mass or
luminosity. Although we have presently no means of testing this
indicator at high redshift, the local relation does not lend much support
to the suggestion that the observed \dv--metallicity relation at high
redshift can be interpreted as a mass--metallicity relation.

\section{Interpretation}
\label{interpretation.sec}
Modeling of DLA velocity profiles already suggested that high
redshift DLAs cannot simply arise in rotationally supported cold gas disks
such as those observed in local galaxies \citep{Prochaska1997a,Haehnelt1998a}. "In this paper we have demonstrated empirically that
the typical velocity profiles
of $z=0$ \hi\ gas above the DLA limit have lower widths than those
seen in DLAs. This result is fully consistent with that of
\citet{Prochaska2002b}, based on the analysis of a single low mass galaxy. How
can these large velocity widths in DLAs be interpreted?

It is well established that most actively star-forming 
high-redshift galaxies show
large-scale outflows, with velocities of several 100 \kms\ up to 1000
\kms. Several authors have suggested that these outflows could be
related to a large fraction of the DLA cross-section at these
redshifts \citep{Nulsen1998a, Schaye2001a}.  \citet{Nulsen1998a}
argued that the starburst accompanying the formation of dwarf
galaxies blows out most of the available gas, which results in a
weakly collimated wind. Once the starburst has subsided, the
(largely ionized) ejected gas recombines and shows \hi\ column
densities typical of DLAs. \citet{Schaye2001a} built on these results
and showed that the incidence rate of DLAs is in agreement with the
wind scenario. He concluded that both disk gas and winds contribute to
the high redshift DLAs cross section.

\subsection{\hi\ shells}
At $z=0$, less extreme examples of star-formation driven blow-out can be
found in \hi\ shells, which are ubiquitous in high resolution
21-cm maps of nearby galaxies \citep[see e.g.][]{Brinks1986a, Puche1992a, Walter1999a, Kim1999a}.
Radiative and kinematic pressure of stellar winds and supernovae from young massive 
stars in OB associations  are thought to lie at the origin of these expanding shells.
\citet{Prochaska2005a} looked specifically at the issue of whether
\dv\ measurements are affected by these shells in the 30 Doradus region in the LMC.
It was found that velocity widths of sight-lines through the \hi\ bubbles  
were only enhanced by approximately 15 \kms\ with respect to the average LMC value.
This result may not come as a surprise, because the typical expansion velocities of the
shells range from 10 to 30 \kms\ in the LMC \citep{Kim1999a}. Similar numbers are found for other nearby galaxies \citep[e.g.,][]{Walter1999a}. Not only is the velocity width small
in shell regions, the area filling factor of \hi\ shells is also small. \citet{Prochaska2002b} quote a
covering factor of $\approx 20$\% for regions above the DLA \nhi\ limit. 

If the physics of supernova feedback  at high redshifts is similar to what is seen in local galaxies, the above results imply that \hi\ shells are unlikely to explain the high velocity widths observed in DLAs. In order to explain the kinematics in high-$z$ DLAs with outflows, one has to invoke more extreme outflow scenarios, such as that seen in star-burst galaxies.

\subsection{Superwinds and tidal features}
To investigate this further, we focus on M82, the prototypical starburst galaxy in the nearby
universe. This galaxy displays a well-documented example of the superwind phenomenon, driven by the molecular driven central starburst, which is  very prominent in X-ray and H$\alpha$ emission \citep[e.g.,][]{Martin1998a, Lehnert1999a}.  M82 is also a member of the M81 group of galaxies and experiences a strong tidal interaction with its neighbors M81 and NGC 3077 \citep{Yun1994a, Walter2002a}. This interaction results in several \hi\ tidal features in the close vicinity of M82. At a distance of only $D=3.6$ Mpc, the galaxy can be observed in great detail, and therefore has been the object of studies at wavelengths ranging from the X-rays to the radio \citep[see][for references]{Engelbracht2006a}. Here, we investigate the \hi\ distribution in M82, making use of the original C-array and D-array Very Large Array (VLA) data of \citet{Yun1993a}, augmented  with B-array data from the VLA archive. The final \hi\ column density map is very similar to the one published in \citet{Taylor2001a}.

\begin{figure}
\begin{center}
\includegraphics[angle=270,width=8.7cm]{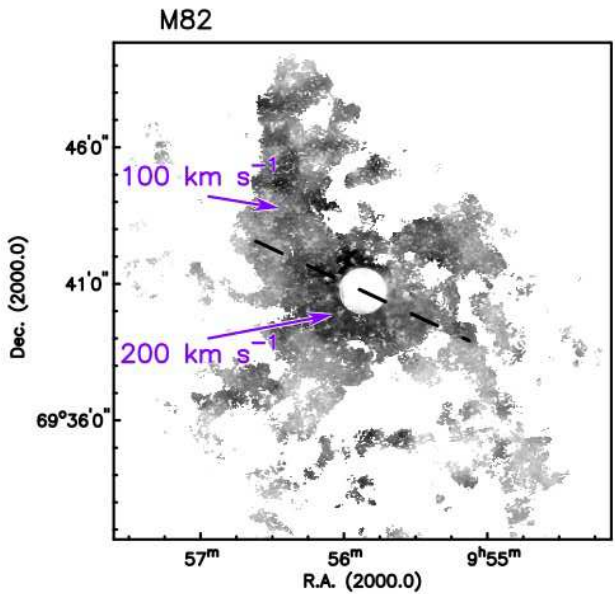}
\caption{The distribution of \dv\ values in M82. Only the regions above the 
DLA column density limit are shown. The central region is blanked because of the
strong radio continuum source. The full range of the greyscale is 0 to 200 \kms.
The dashed line indicates the major axis of the
stellar distribution in M82. Regions of $\dv\approx 100$ \kms\ and $\dv\approx 200$ \kms\
are indicated in the north-east tail and in the central regions, respectively.
\label{M82.fig}}
\end{center}
\end{figure}

We use the same procedure as described for the THINGS galaxies to calculate \dv\ values from the M82 data  for the regions of the \hi\ gas above the DLA column density limit. Figure \ref{M82.fig} shows the distribution of \dv\ in M82. We find that the median \dv\ from these data is $\approx 110$ \kms. Very high values of $\dv\approx 200$ \kms\ are seen in the central region over an area of several kpc$^2$. 
Along the minor axis of M82 these high velocity widths are attributed to interactions of M82's gaseous medium with the superwind \citep{Yun1993a}. Interestingly, similarly high velocity widths along the minor axis are observed in CO \citep{Taylor2001a,Walter2002a}. 
The tidal features in M82 also show high velocity widths. The south-west tail that connects up to M81 is not well seen in our map, but the north-east tail shows \dv\ values 
around 100 \kms. 

Obviously, M82's high inclination to the line-of-sight ($\approx 80\deg$) will contribute to the high median \dv\ observed in this galaxy. However, the gaseous regions in this galaxy that are clearly identified as either  tidal features or those related to the superwind display similarly high velocity spreads. As we investigated only a single starburst galaxy it is difficult to draw quantitative conclusions on the basis of these data. However, we argue that M82 clearly demonstrates that the superwind phenomenon and tidal interactions, which are both more frequent at higher redshifts, could go a long way in explaining the high \dv\ values observed in DLAs. Both winds and tidal features could contribute to the main peak of the DLA distribution in Figure \ref{histV90.fig}, as well as to the high \dv\ bump.

In this context it is interesting to mention the results of \citet{Bouche2006b}, who find that the equivalent width of MgII absorbers is anti-correlated with the mass of the halo that hosts them. These authors interpret this as evidence that the MgII absorbers are not virialized in the gaseous haloes of the galaxies, but, instead, are produced by superwinds. A fair fraction of these MgII systems are DLAs. \citet{Rao2006a} showed that 35\% of the MgII systems with EW (MgII $\lambda 2796$, FeII $\lambda 2600$) $>$ 0.5 \AA\ have \hi\ column densities above the DLA threshold of $2\times 10^{20}\,\icmsq$.

Additional evidence for a contribution to the DLA cross-section from tidal gas
in galaxy groups like M81, comes from the recent observations of a much smaller sample of strong MgII systems by \citet{Nestor2007a}.
These authors find that these very strong intermediate redshift MgII absorption systems often arise in pairs or small groups of galaxies. An estimated 60\% of these systems are DLAs. 
\citet{Nestor2007a} argues that the likely origin of the high equivalent width MgII absorption is  kinematically disturbed gas around interacting galaxies.

The different peaks in the distribution of velocity width in the low and
high redshift samples in Figure \ref{histV90.fig} could be interpreted as being due to an
increase in the number of interacting systems with redshift. \citet{Gottlober2001a}
use N-boby simulations to show that the number of interacting
systems (i.e. major mergers) scales as $(1+z)^3$. Thus, we would expect a
significant increase in lines-of-sight with large velocity widths at
higher redshifts.

\subsection{High velocity clouds}
\citet{Haehnelt1998a} have argued that irregular protogalactic clumps could (partly) produce 
the large velocity widths of DLAs at high redshift. Perhaps the most closely related examples of such clouds in the nearby universe are Galactic high velocity clouds \citep[HVCs,][]{Wakker1991a}. The \hi\ column densities in the clouds themselves very rarely exceed the DLA limit, but it is conceivable that along a sight-line through the gas disk, HVCs could contribute sufficiently to the integrated optical depth to influence the \dv\ measurements. Several authors have noted the similarity between the properties of Galactic HVCs and those of MgII absorption systems \citep[e.g.][]{Richter2005a,Bouche2006b}, of which DLAs are a subset.

Because of the unknown distances to the vast majority of the Galactic HVCs, their physical 
properties are difficult to ascertain. In this respect, the recent identification of HVCs around M31 is particularly interesting \citep{Thilker2004a}. Synthesis folow-up observations by \citet{Westmeier2005a} show that these clouds have typical sizes of 1 kpc and column densities in the range $10^{19}$ to $10^{20}$ \icmsq, i.e., below the DLA threshold. If these HVCs really are the 
relics of galaxy formation, it is likely that they present a much higher cross-section at higher redshifts when galaxy formation was more actively taking place. As such, they could have a significant effect on the \dv\ measurements in high-$z$ DLAs. We note that this component is closely related to  (and perhaps partly a variant of) the tidal streams and winds discussed above.  

Also interesting in this context are the recent very deep 21-cm observations of 
NGC 2403 \citep{Fraternali2002a} and NGC 891 \citep{Oosterloo2007a}. These observations reveal low column density extended \hi\ haloes, which typically contain 10 to 30\% of the total \hi\ mass of the galaxies\footnote{Related to these observations are the results of \citet{Heald2006a} who show evidence for diffuse ionized extraplanar gas, decoupled from the main rotation of the disk. However, analogs of this ionized gas in DLAs at higher $z$ are unlikely to add much to the velocity widths measured in the low-ionization lines, for reasons explained in section \ref{lowion.sec}}. Again, the \hi\ column densities in the haloes are too low to cause
DLA absorption by themselves, but along  sight-lines through the gas disks, the haloes can have a significant effect on \dv\ measurements. 
These haloes show structures such as clouds and filaments, which resemble the properties of Galactic HVCs. Detailed modeling of the \hi\ haloes \citep{Fraternali2006a} shows that their origin lies partly in a fountain mechanism \citep{Shapiro1976a}, as demonstrated by the close link between the extra-planar \hi\ and H$\alpha$ emission \citep[][and references therein]{Boomsma2007a}. The driving force behind this fountain is star formation in the disk. The cosmic star formation rate density was much higher in the past, up to a factor 20 higher at $z=2-3$ compared to the present epoch \citep{Hopkins2005a}. The increase in the cosmic neutral hydrogen mass density is much less dramatic, approximately a factor 2 over same redshifts range \citep{Prochaska2005a}. These two results combined imply that the energy input
into the cold gaseous medium per unit gas mass was much higher in the past. This may lead to a relatively high contribution of fountain gas to the kinematics of cold gas at these redshifts. This high velocity gas would probably be characterized by profiles very similar to the ones observed in DLAs displaying a main peak and a weaker tail.
In passing, we note that the higher energy input into the gas disk would also lead to an increase in the gas turbulence, which also causes larger velocity widths. 

In order to fully explain the properties of the gaseous haloes,  \citet{Fraternali2006a} speculate that accretion from the surrounding intergalactic medium is also required. This ``cold mode" of gas accretion is increasingly more important at higher redshifts in the smoothed particle hydrodynamics simulations of \citet{Keres2005a}, indicating that this halo gas could play an important role in the gas kinematics of DLAs.

%################
\section{Conclusions}
\label{conclusions.sec}

In this paper we have tested whether the kinematics of damped Ly$\alpha$ absorbers at high redshift are consistent with the idea that these systems arise in the gas disks of galaxies such as those in the local universe. We have examined the gas kinematics of local galaxies using the high quality \hi\ 21-cm data from the \hi\ Nearby Galaxies Survey (THINGS). To characterize the velocity widths, we adopt the parameter \dv, the width that encompasses the central 90\% of the total flux seen along any sightline through the galaxies. The same parameter is used for DLAs to measure the width of the low-ionization lines, which are believed to trace the kinematics of the neutral gas.

We can use the $z=0$ THINGS data to calculate in detail how \dv\ depends on \hi\ column density, and we find that larger \dv\ values are more likely to occur at low \hi\ column densities. 
A similar (but weak) trend is seen for the DLAs, albeit at at higher values of \dv. 
We also investigate how \dv\ depends on 
impact parameter from the center of a galaxy and on galaxy luminosity. However, at 
present it is difficult to make a quantitative  comparison between these calculations and the DLA data because of the very few identifications of DLA galaxies in the literature. For a proper
understanding of the kinematics and other properties of DLAs, it is important that more DLA galaxies be identified at low and high redshifts.

The most important result is that the median \dv\ at $z=0$ is approximately 30 \kms, a factor of two lower than what is seen in DLAs at redshifts $z\sim 1.5-4$.  
We discuss several processes that could increase  the velocity widths in DLAs, and conclude that tidal tails, superwinds, accretion, as well as fountain gas are all likely to contribute to the kinematics of DLAs. Each of these processes are enhanced at higher redshifts, where galaxy interactions are more common, superwinds are more frequently observed, `cold mode' accretion is more efficient, and the star formation rate density is higher.

\acknowledgments
We thank N. Bouch\'e and C. P\'eroux for their comments and the anonymous referee for a constructive report, which helped improving the presentation of this paper.
J. Ott and A. Weiss  are thanked for their help with the reduction of 
the M82 VLA data used in this analysis. EB gratefully acknowledges financial support through an EU Marie Curie
International Reintegration Grant (Contract No. MIRG-CT-6-2005-013556. The National Radio Astronomy Observatory is a facility of the National Science Foundation operated under cooperative agreement by Associated Universities, Inc. 

\bibliographystyle{apj}

\clearpage

\end{document}